\documentclass[useAMS,usenatbib]{mn2e}

\usepackage{graphicx}
\usepackage{hyperref}
\usepackage{enumerate}
\usepackage{xcolor}
\usepackage{longtable}
\usepackage{txfonts}
\usepackage{times}

\newcommand{\hi}{{\sc H\,i}}

\newcommand{\ltsima} {$\; \buildrel < \over \sim \;$}
\newcommand{\gtsima} {$\; \buildrel > \over \sim \;$}
\newcommand{\lta} {\lower.5ex\hbox{\ltsima}}
\newcommand{\gta} {\lower.5ex\hbox{\gtsima}}

\newcommand{\atlas}{ATLAS$^{\rm 3D}$}

\newcommand{\sofia}{\texttt{SoFiA}}
\newcommand{\duchamp}{\texttt{Duchamp}}

\defcitealias{cappellari2011a}{Paper I}
\defcitealias{krajnovic2011}{Paper II}
\defcitealias{emsellem2011}{Paper III}
\defcitealias{young2011}{Paper IV}
\defcitealias{davis2011a}{Paper V}
\defcitealias{bois2011}{Paper VI}
\defcitealias{cappellari2011b}{Paper VII}
\defcitealias{khochfar2011}{Paper VIII}
\defcitealias{duc2011}{Paper IX}
\defcitealias{davis2011b}{Paper X}
\defcitealias{serra2012a}{Paper XIII}
\defcitealias{cappellari2013a}{Paper XV}
\defcitealias{krajnovic2013a}{Paper XVII}
\defcitealias{cappellari2013b}{Paper XX}
\defcitealias{krajnovic2013b}{Paper XXIII}
\defcitealias{naab2013}{Paper XXV}

\title[SoFiA 3D source finder]{SoFiA: a flexible source finder for 3D spectral line data}
\author
[Paolo Serra et al.]{\parbox{\textwidth}{Paolo Serra,$^{1}$\thanks{E-mail: \texttt{paolo.serra@csiro.au}}
Tobias Westmeier,$^2$
Nadine Giese,$^3$
Russell Jurek,$^1$
Lars Fl\"{o}er,$^4$
Attila Popping,$^{2,5}$
Benjamin Winkel,$^6$
Thijs van der Hulst,$^3$
Martin Meyer,$^2$
B\"{a}rbel S. Koribalski,$^1$
Lister Staveley-Smith$^{2,5}$
and H\'{e}l\`{e}ne Courtois$^{7}$
}
\vspace{0.4cm}\\ 
\parbox{\textwidth}{$^{1}$CSIRO Astronomy and Space Science, Australia Telescope National Facility, PO Box 76, Epping, NSW 1710, Australia\\
$^{2}$ICRAR, M468, The University of Western Australia, 35 Stirling Highway, Crawley, WA 6009, Australia\\
$^{3}$University of Groningen, Kapteyn Astronomical Institute, Landleven 12, NL-9747 AD, Groningen, the Netherlands \\
$^{4}$Argelander-Institut f\"{u}r Astronomie, Auf dem H\"{u}gel 71, 53121 Bonn, Germany\\
$^5$ARC Centre of Excellence for All-sky Astrophysics (CAASTRO)\\
$^{6}$Max-Planck-Institut f\"{u}r Radioastronomie, Auf dem H\"{u}gel 69, 53121 Bonn, Germany\\
$^{7}$Universit\'{e} Lyon 1, CNRS/IN2P3, Institut de Physique Nucl\'{e}aire, Lyon, France\\
}}

\begin{document}

\date{Accepted 2014 December 30.  Received 2014 December 22; in original form 2014 July 28}

\pagerange{\pageref{firstpage}--\pageref{lastpage}} \pubyear{2013}

\maketitle

\label{firstpage}


\begin{abstract}
We introduce \sofia, a flexible software application for the detection and parameterization of sources in 3D spectral-line datasets. \sofia\ combines for the first time in a single piece of software a set of new source-finding and parameterization algorithms developed on the way to future \hi\ surveys with ASKAP (WALLABY, DINGO) and APERTIF. It is designed to enable the general use of these new algorithms by the community on a broad range of datasets. The key advantages of \sofia\ are the ability to: search for line emission on multiple scales to detect 3D sources in a complete and reliable way, taking into account noise level variations and the presence of artefacts in a data cube; estimate the reliability of individual detections; look for signal in arbitrarily large data cubes using a catalogue of 3D coordinates as a prior; provide a wide range of source parameters and output products which facilitate further analysis by the user. We highlight the modularity of \sofia, which makes it a flexible package allowing users to select and apply only the algorithms useful for their data and science questions. This modularity makes it also possible to easily expand \sofia\ in order to include additional methods as they become available. The full \sofia\ distribution, including a dedicated graphical user interface, is publicly available for download.
\end{abstract}

\begin{keywords}
methods: data analysis.
\end{keywords}

\begin{figure}
\includegraphics[width=8.45cm]{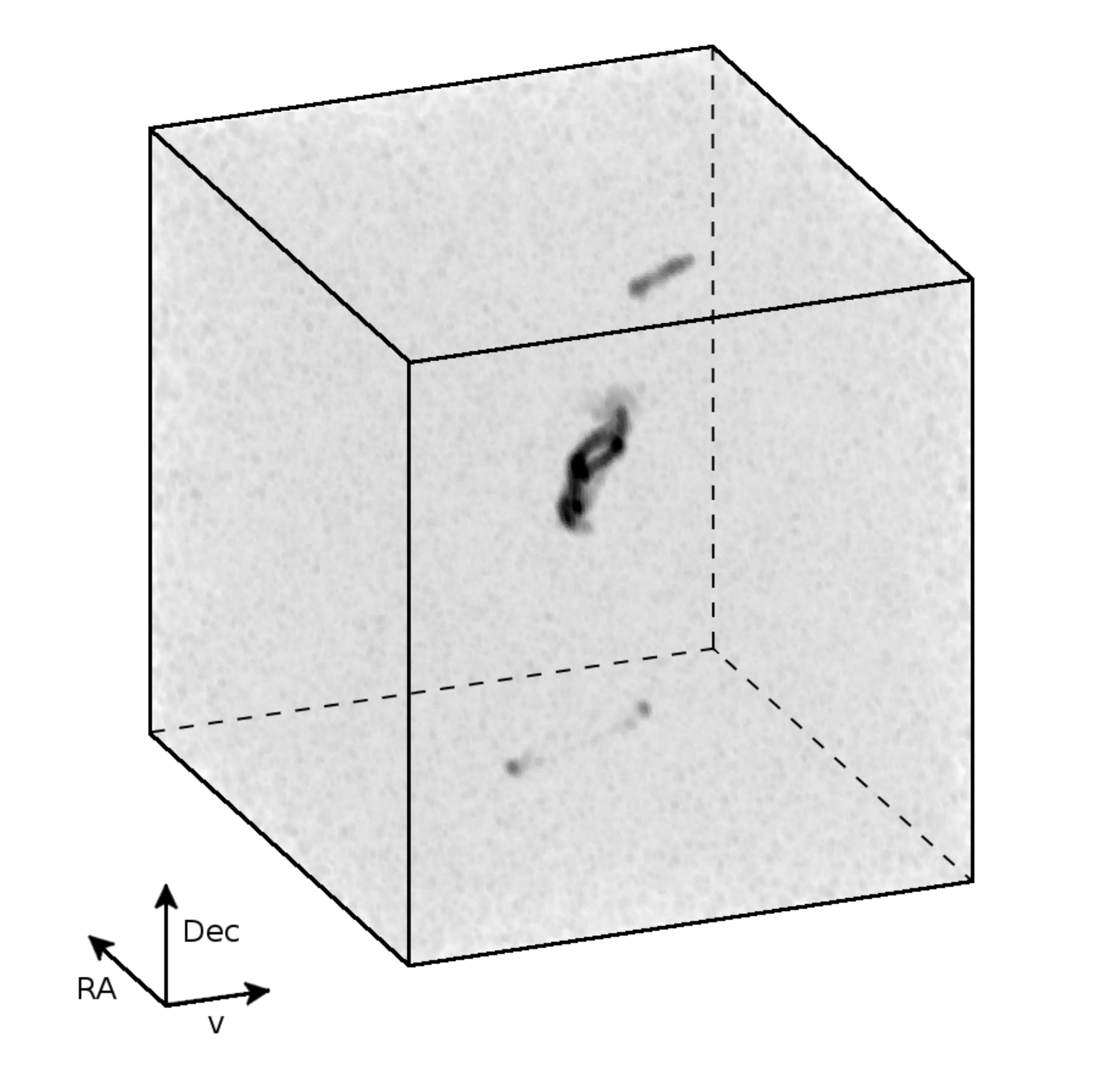}
\caption{Volume rendering of an \hi\ data cube showing that individual sources have a complex and diverse 3D structure. This makes their detection and accurate parameterization challenging.}
\label{fig:paperfiguredavide}
\end{figure}

\begin{figure}
\includegraphics[width=8.41cm]{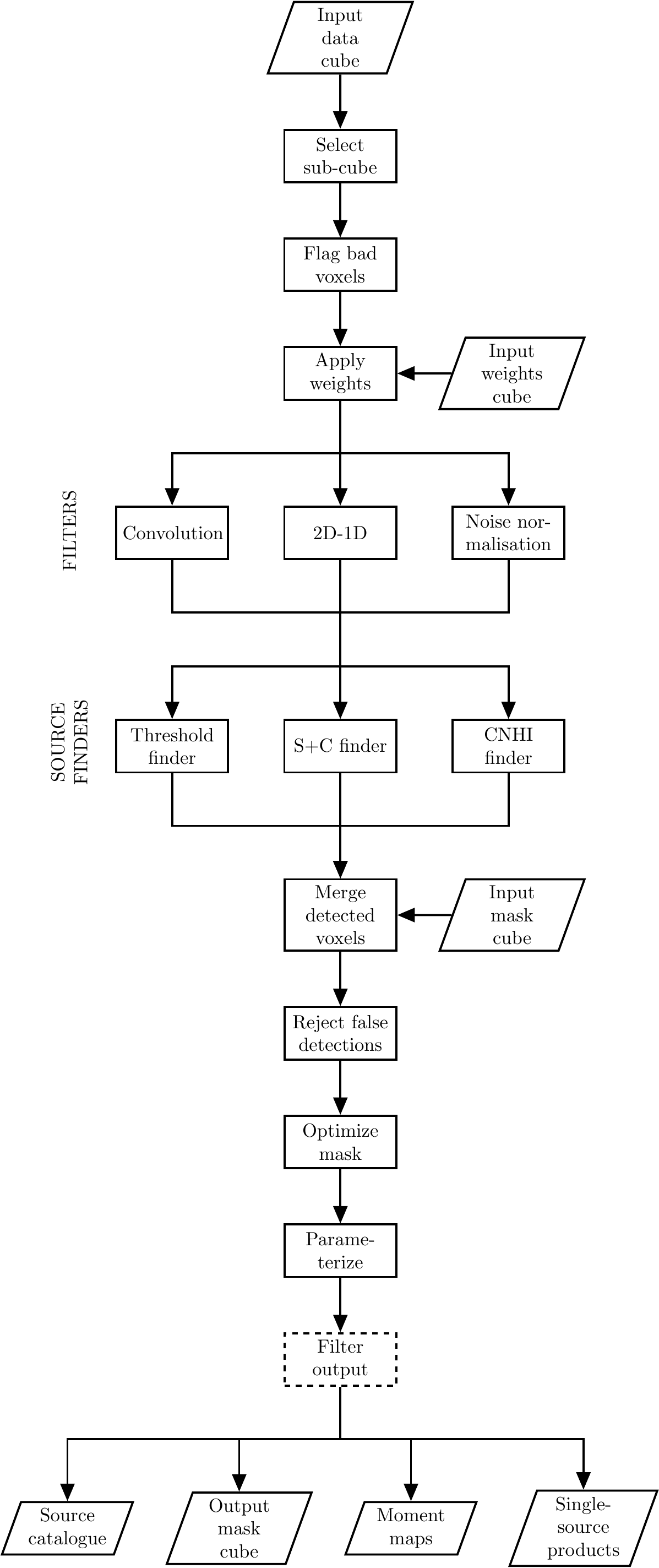}
\caption{\sofia\ flowchart. We highlight the ``Filter output'' module with a dashed box as this will become available in future releases of \sofia.}
\label{fig:flowchart}
\end{figure}

\section{Introduction}
\label{sec:intro}

The detection of astronomical signal above instrumental noise is a crucial aspect of all astronomy observations. The techniques employed to detect and characterise this signal depend on the type of data being analysed (see \citealt{masias2012} for a review). Standard methods and tools have emerged in fields with a large community base such as 2D imaging (e.g., \texttt{SExtractor}; \citealt{bertin1996}) and 1D spectroscopy (e.g., \texttt{GANDALF}; \citealt{sarzi2006}). In other fields with relatively fewer users, detection algorithms vary significantly between projects. This is the case for studies based on 3D spectral line data (for brevity, data cubes), where the flux of a spectral line is mapped as a function of position on the sky and line-of-sight velocity of the emitting matter.

The diversity of source finding methods for data cubes is at least partly due to the diversity of 3D structure of the sources being studied. We illustrate this point in Fig. \ref{fig:paperfiguredavide}, where we show a data cube of the \atlas\ \hi\ survey \citep{serra2012a}. In this figure, the central object is bright (and therefore easy to detect) but has a complex 3D structure, including a low surface brightness extension towards large RA. On the contrary, the top object is bright and relatively simple as emission is confined within a small range of RA and Dec. Finally, the bottom object is the typical case of a resolved, edge-on galaxy where the two peaks of the double-horn velocity profile are clearly visible, and detection of the faint emission between the two peaks is challenging. An ideal 3D source finder should be able to detect and parameterize all these different sources in a complete and reliable way.

Radio single dishes and interferometers have traditionally been the most common telescopes used to construct data cubes (although optical integral-field spectrographs are now also generating large numbers of such cubes -- e.g., \citealt{cappellari2011a,croom2012,sanchez2012}). The upgrade and continuing operation of existing radio telescopes, as well as the construction of the Square Kilometre Array and its precursors, are leading to a rapid increase in the number and size of data cubes. Standard and sufficiently general source-finding tools will be necessary to analyse these data, and recent work has started addressing this need (see, e.g., \duchamp\ by \citealt{whiting2012a}). In this paper we introduce \sofia, a new, flexible Source Finding Application for data cubes which combines detection algorithms and techniques from several source finders.

\sofia\ is designed to work on any data cube independent of telescope or observed spectral line. However, its development is part of preparatory work for a few specific, upcoming \hi\ surveys: WALLABY, a blind \hi\ survey of 3/4 of the entire sky out to $z\sim0.25$ to be carried out with the Australian Square Kilometre Array Pathfinder (ASKAP; see \citealt{koribalski2012b}); DINGO, a deep \hi\ survey out to $z\sim0.4$ (also to be carried out with ASKAP; see \citealt{meyer2009}); and the \hi\ surveys planned for APERTIF \citep{verheijen2008}. This preparatory work has resulted in the development of a number of new source-finding algorithms, which are described in a series of papers referred to in the next Section (for a summary see \citealt{koribalski2012a}). \sofia\ puts these different algorithms together for the first time in a coherent, flexible and publicly available piece of software.

\sofia\ can be obtained from \url{https://github.com/SoFiA-Admin/SoFiA} . On the same webpage we provide a list of requirements, installation instructions and a user manual. The aim of this paper is to describe how \sofia\ operates on data cubes and thereby provide a reference for current and future users.

\section{Description of \sofia}

\sofia\ is a modular application whose aim is to detect and parameterize sources in a data cube. The flowchart in Fig. \ref{fig:flowchart} shows the various modules that users can choose to use (or not to use), in the order in which they are executed by \sofia. Once an input data cube (or a sub-cube selected by the user) is loaded, these modules allow users to:
\begin{itemize}
\item{modify the input cube by applying flags, weights, or a set of filters;}
\item{detect the spectral line signal;}
\item{identify sources by merging detected voxels together;}
\item{reject false detections;}
\item{optimize the mask of individual sources;}
\item{measure source parameters;}
\item{filter the output by selecting a region of interest in source parameter space;}
\item{and produce output catalogues as well as cubes, moment maps, position-velocity diagrams and integrated spectra.}
\end{itemize}
Individual modules are described in more detail in the rest of this Section. They are written in either Python or C++ and rely on a range of external libraries, including NumPy and SciPy \citep{scipy-collaboration,Walt2011}, Cython \citep{behnel2011}, Astropy \citep{2013A&A...558A..33A}, the GNU Scientific Library\footnote{\url{http://www.gnu.org/software/gsl/}} and, optionally, matplotlib \citep{Hunter2007}. Provided that these libraries are available, \sofia\ can run on all machines with a Unix or Linux operating system (including, e.g., Mac~OS~X and Ubuntu). We refer to the \sofia\ webpage for up-to-date details.

\begin{figure}
\includegraphics[width=8.45cm]{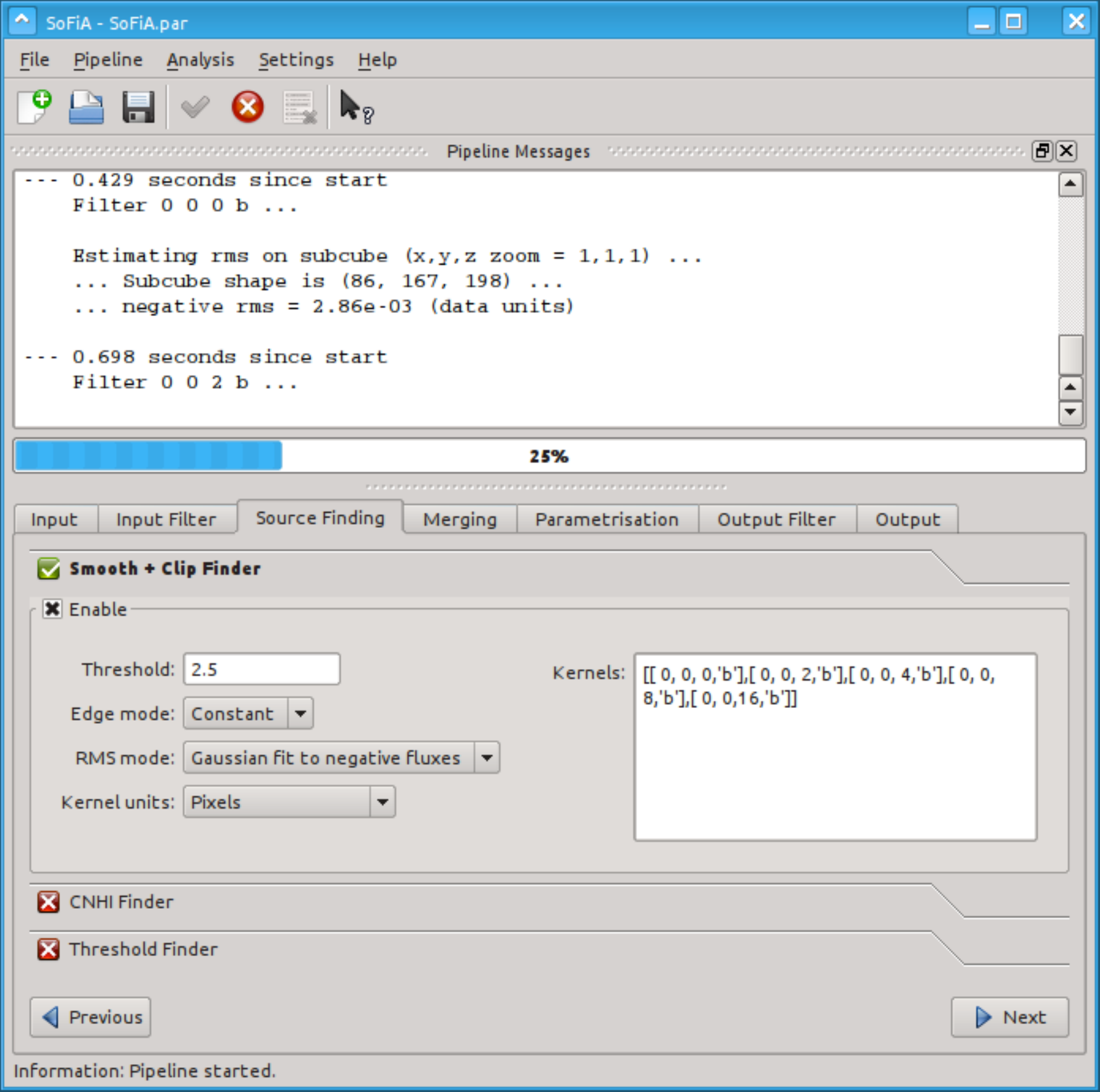}
\caption{Screenshot of the \sofia\ GUI. The GUI adopts automatically the native style of the window manager used on the system where \sofia\ is installed. In this figure we show the GUI as it appears on a Kubuntu Linux system. The GUI also offers the option of displaying the source catalogue generated by \sofia\ and includes a help browser that explains the available parameter settings.}
\label{fig:paperfigure0}
\end{figure}

\sofia\ can be executed from the command line or using a dedicated graphical user interface (GUI) based on the Qt library (see Fig. \ref{fig:paperfigure0}). Both methods allow users to select which combination of the above modules and which source finding and parameterization algorithms to use. This selection is done using either the GUI or a plain text parameter file (if running \sofia\ from the command line), allowing the source-finding strategy and its complexity to be optimized for the type of data and sources of interest. For example, \sofia\ could be asked the simple question of creating a moment-0 image of all voxels above a given threshold in a data cube -- in which case most of \sofia's functionalities would be switched off. Alternatively, it could be given a number of relatively more complex tasks such as, for example, applying a wavelet filtering algorithm, rejecting false detections or fitting models to the spectrum of the detected sources.

While \sofia\ will continue to be improved, this basic principle of modularity will not change. Therefore, although this paper describes the software as it is at the time of writing and new algorithms may be introduced in the future, the main workings of \sofia\ will remain as illustrated here.

\subsection{Data cube, weights cube, mask cubes and filters}
\label{sec:cubes}

Four different types of input and/or output cubes are relevant at different stages of \sofia.
\begin{itemize}
\item \it Data cube\rm, which includes signal from astronomical sources superimposed on instrumental noise (and errors).
\item{\it Weights cube\rm, which allows users to weight voxel values to take into account, e.g., noise level variations across the cube or the presence of imaging artefacts in certain regions of the data cube.}
\item{\it Binary mask\rm, where detected and non-detected voxels have values of 1 and 0, respectively.}
\item{\it Object mask\rm, where non-detected voxels have a value of 0 and detected voxels have an integer value corresponding to the ID of the object they belong to.}
\end{itemize}

All source-finding algorithms implemented in \sofia\ and described in Sec. \ref{sec:det} below assume that the noise level is uniform across the data cube. Therefore, noise variations caused by, e.g., mosaicking or frequency-dependent flagging need to be removed first. This can be done within \sofia\ by means of a weights cube inversely proportional to the noise level. \sofia\ removes noise variations by multiplying the data cube by the weights cube. Once source detection is completed, \sofia\ will undo this operation before measuring source parameters. The weights cube could also be useful to down-weight regions of a data cube affected by imaging artefacts (e.g., cleaning or continuum-subtraction residuals).

The weights cube can be provided by the user. Alternatively, users can provide an analytic description of the weights variation across the cube. Finally, a weights cube inversely proportional to the local noise level can be derived by \sofia\ and applied to the data cube. The evaluation of the local noise level is carried out independently along any or all of the three axes of the data cube. For example, a user may wish to remove noise variations along the frequency axis alone, under the assumption that the noise does not vary within each frequency plane.

We note that \sofia\ measures the noise within a data cube at various other stages of the processing. Different methods of noise measurement are implemented and users can decide which one is more appropriate for their purpose. Possible choices are: \it i) \rm standard deviation; \it ii) \rm median absolute deviation; and \it iii) \rm standard deviation of a zero-centred Gaussian fit to the negative side of the flux histogram.

The calculation and application of the inverse-noise weights cube described above is part of a more general SoFiA module which allows users to apply a filter to the data cube before running the selected source-finding and parameterization algorithms. As indicated in Fig. \ref{fig:flowchart}, this module includes two additional filtering methods: firstly, the convolution with a 3D kernel whose shape can be chosen among a few options and whose size can be specified by the user; and secondly, the 2D-1D wavelet de-noising algorithm developed by \cite{floer2012}. This algorithm processes the two spatial dimensions and the spectral dimension of the data cube separately, and returns a noise-free data cube reconstructed using only wavelet coefficients above a specified threshold. Additional filtering options may be provided in future releases.

As indicated by the flowchart in Fig. \ref{fig:flowchart}, portions of the cube can be blanked out (flagged) prior to source finding. This may be necessary at the location of very bright continuum sources whose spectrum was not subtracted properly from the data, or at channels dominated by line emission from the Galaxy or affected by strong radio frequency interference.

Finally, mask cubes are generally calculated within \sofia\ (see below) but can also be provided by the user. The latter could be desirable if a user, following an initial source-finding run, wishes to look for additional sources with a different search algorithm or parameters. In this case the new sources are added to the initial, input mask. Alternatively, an input mask could be used if sources have already been identified and only subsequent parameterization steps are required.


\subsection{Detection of spectral line signal}
\label{sec:det}

\sofia\ is meant to offer a number of detection algorithms that users can choose from. A common advantage of these algorithms is the ability to look for emission on multiple scales, which is essential to detect sources in 3D (see Fig \ref{fig:paperfiguredavide}). An exception is the simple threshold method (see below), unless used in combination with some of the filtering methods described above (e.g., 2D-1D wavelet denoising). The following algorithms are implemented in \sofia.

\begin{itemize}
\item Simple threshold. This is the simplest possible algorithm (and the only one not operating on multiple scales): only voxels whose absolute value is above a specified threshold are detected. Users can specify the threshold in flux units or relative to the noise level.
\item S+C. This is the smooth + clip algorithm developed by \cite{serra2012a} on the basis of techniques traditionally used within the \hi\ community. It consists of searching for emission at multiple angular and velocity resolutions by smoothing the data cube with 3D kernels specified by the user. At each resolution, voxels are detected if their absolute value is above a threshold given by the user (in noise units). The final mask is the union of the masks constructed at the various resolutions.
\item CNHI. This algorithm was developed by \cite{jurek2012}. Individual 1D spectra (or bundles of adjacent spectra) are extracted from the data cube. For each of them, the Kuiper test is used to identify regions of the spectrum which are not consistent with containing only noise. In practice, users need to provide a probability threshold above which a spectral region is considered detected and is added to the final binary mask.
\end{itemize}
The numerous possible combinations of these source-finding methods together with the filtering algorithms described in Sec. \ref{sec:cubes} allow users to design a number of different strategies to detect signal in their data cube. For example, the CNHI finder could be run following convolution with a 3D kernel appropriate for the type of sources being searched. Alternatively, a simple threshold method could be used after the noise has been removed from the cube by the 2D-1D wavelet filter.

\cite{popping2012} discuss strengths and weaknesses of these algorithms and compare their performance. An important recommendation of that work is that all source finders should incorporate some form of 3D smoothing in order to increase completeness. In this respect, the simple threshold algorithm is of limited use unless coupled with a filtering methods such as the 2D-1D wavelet de-noising. \cite{popping2012} find this particular combination to deliver higher completeness and reliability than S+C and CNHI for sources unresolved on the sky, especially at narrow line widths. In contrast, the S+C method is by construction well suited to finding sources on a variety of scales and \cite{popping2012} deem it the best choice for extended objects.

We note that many of the algorithms have been improved since the comparative study of \cite{popping2012}. Additional testing can now be carried out within \sofia\ and will be used to investigate how to further improve their performance. Until then, we refer to the aforementioned papers for a complete discussion of these methods.

\begin{figure*}
\includegraphics[height=4.1cm]{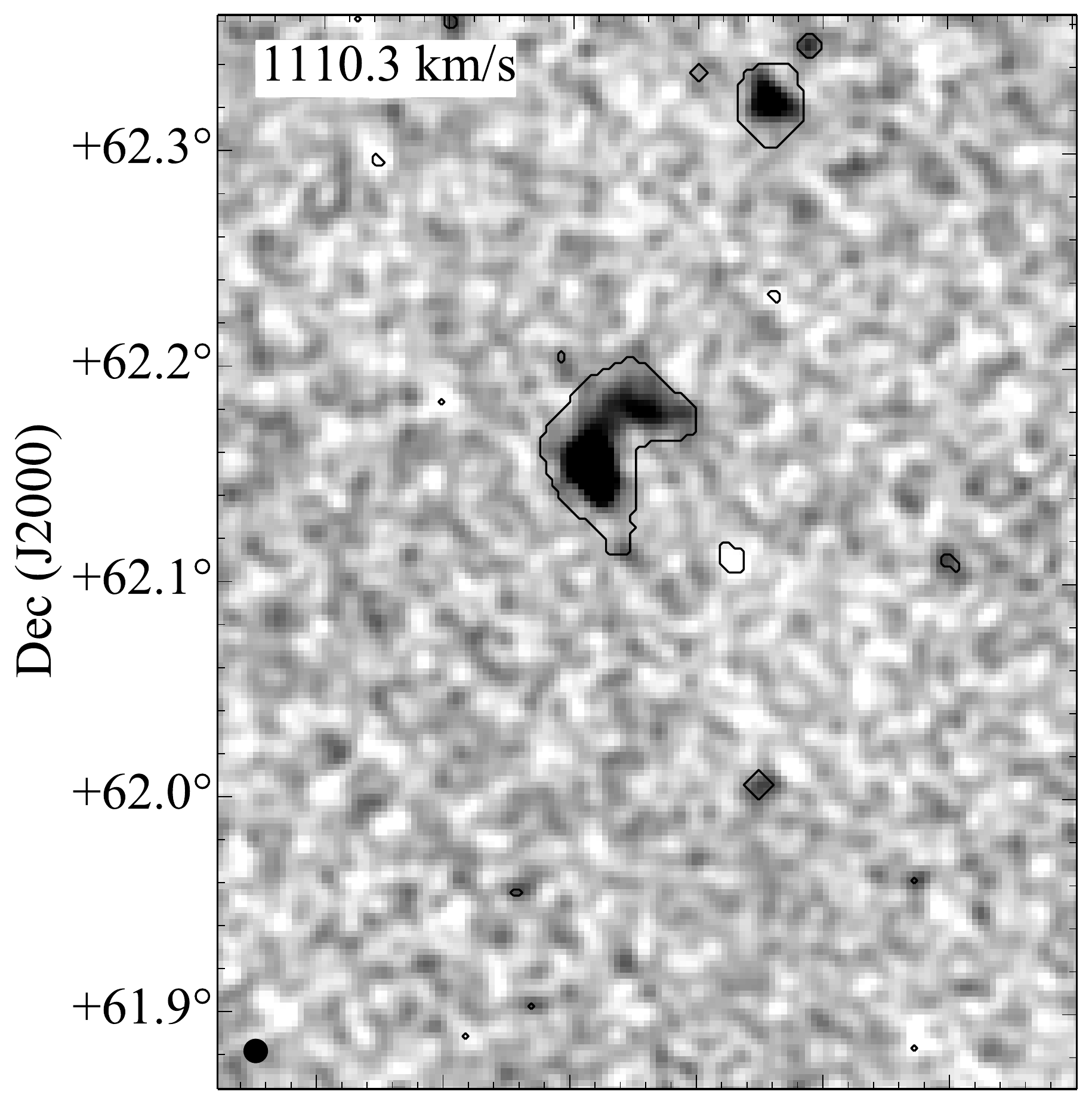}
\includegraphics[height=4.1cm]{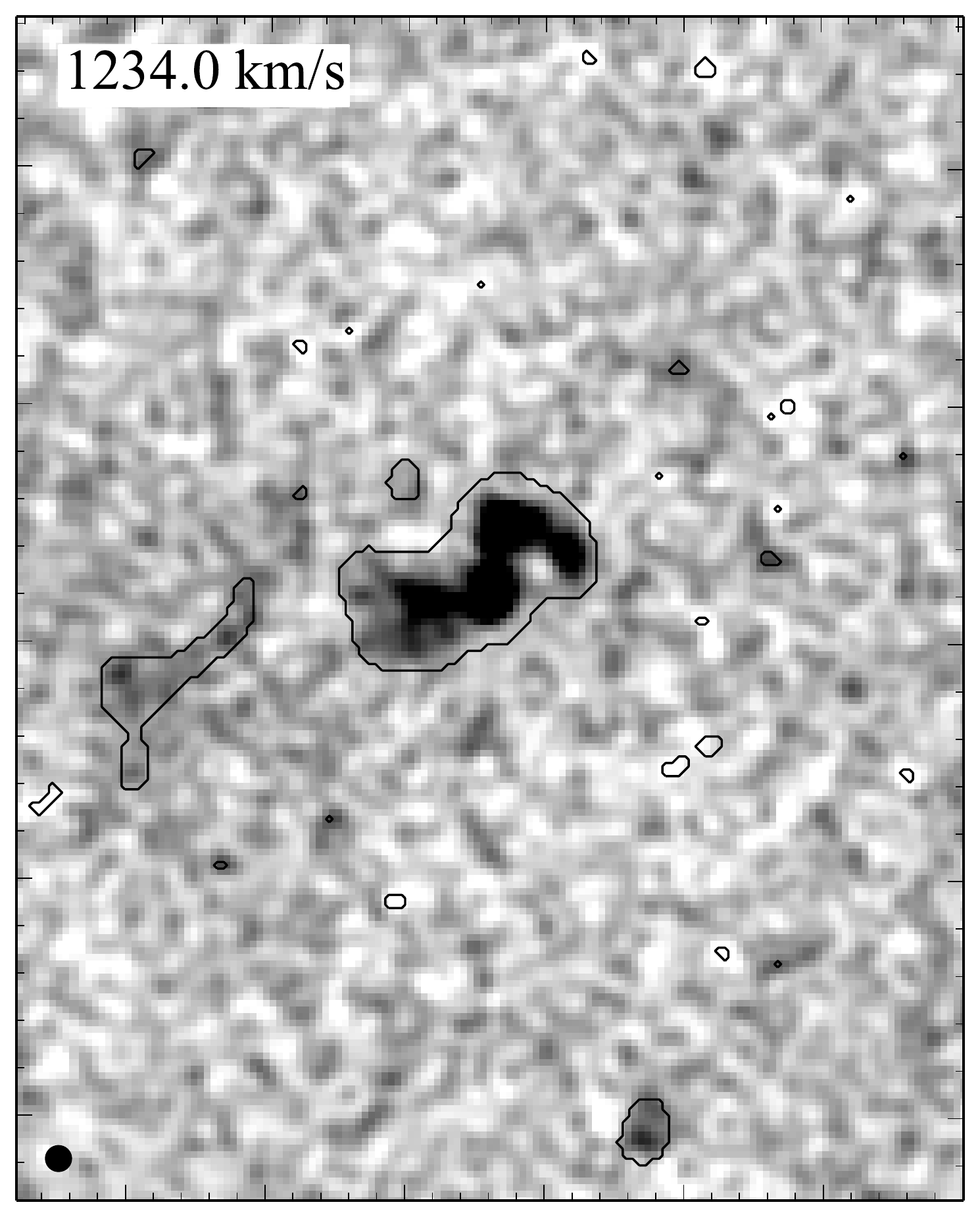}
\includegraphics[height=4.1cm]{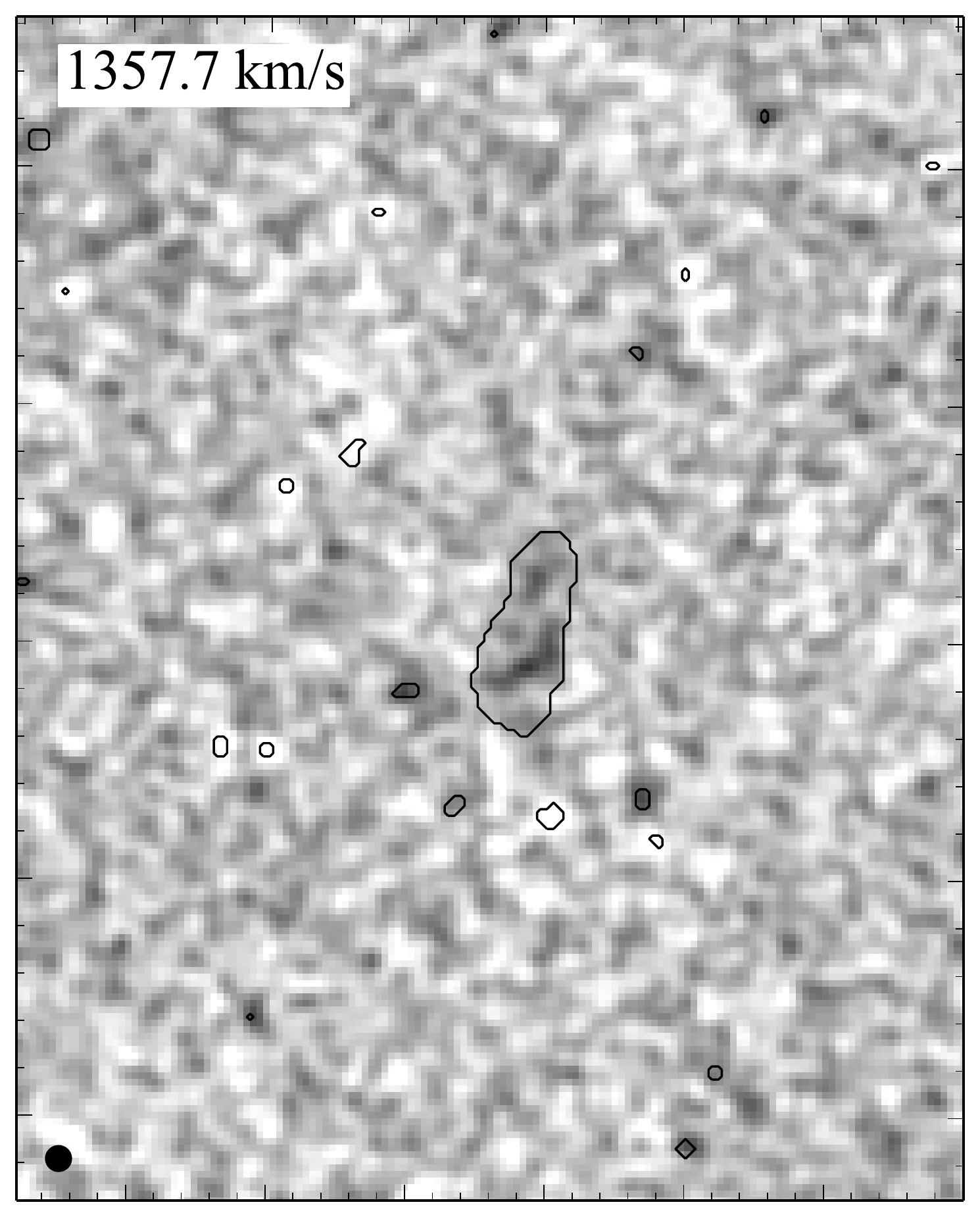}
\includegraphics[height=4.1cm]{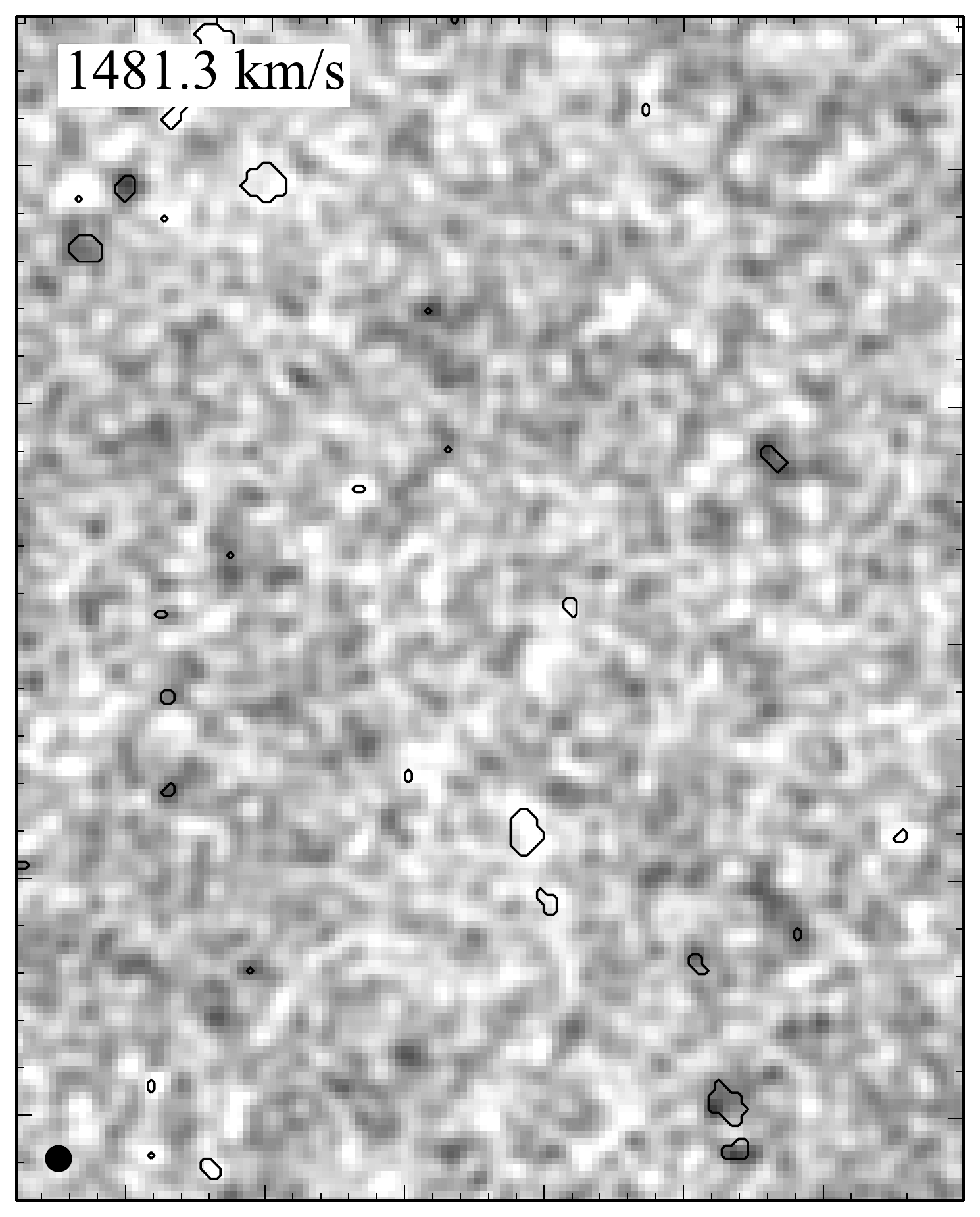}
\includegraphics[height=4.1cm]{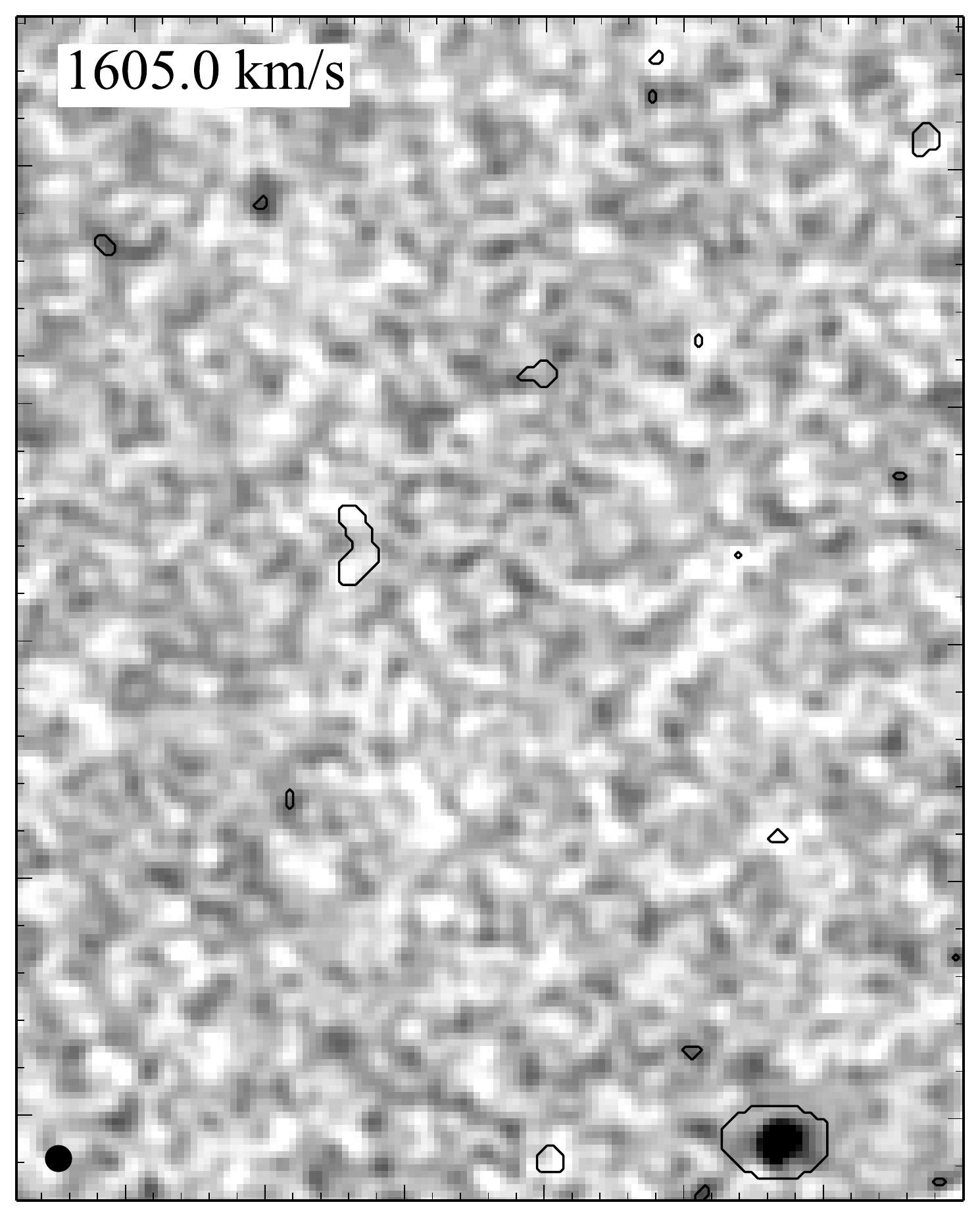}
\includegraphics[height=4.52cm]{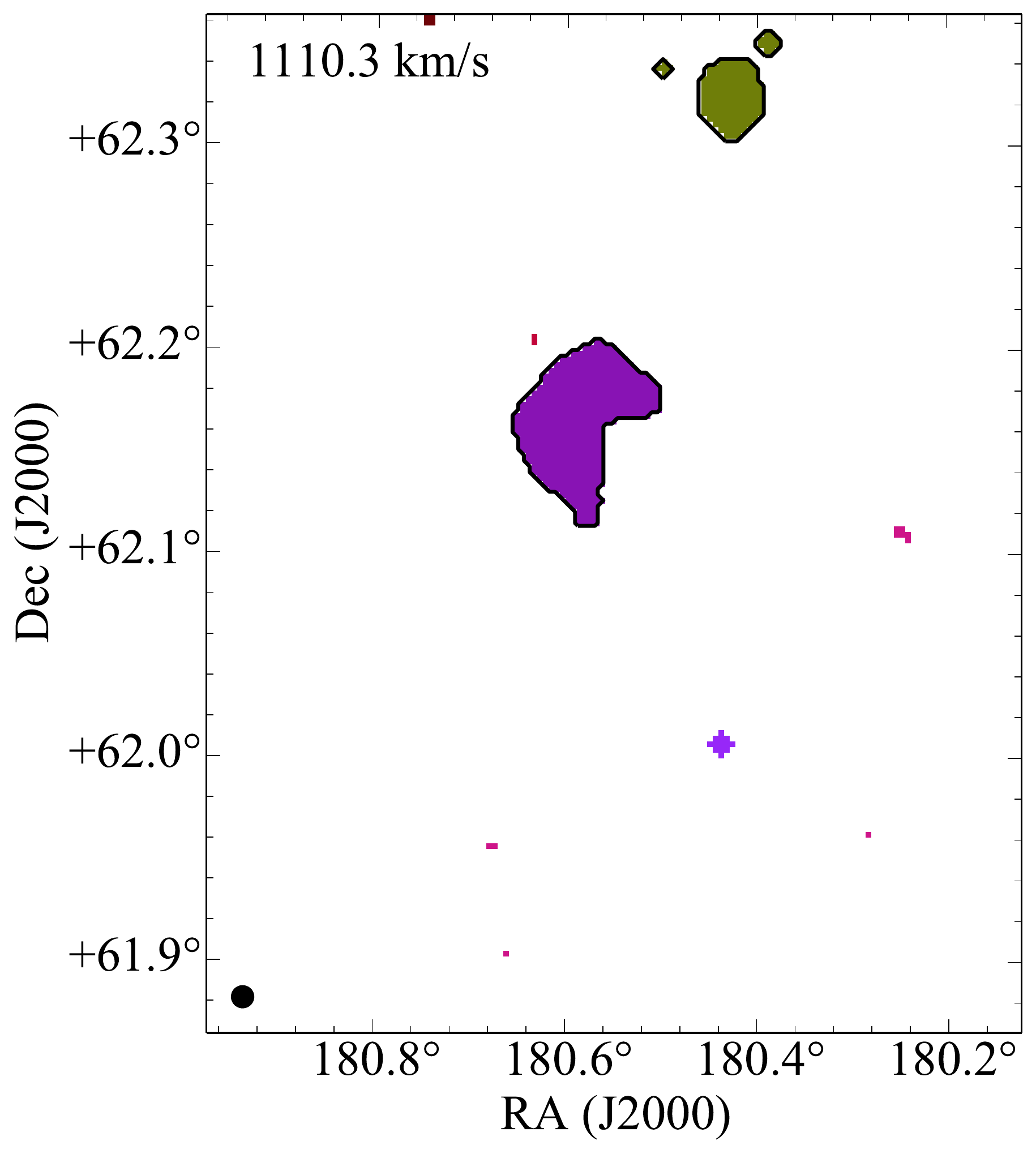}
\includegraphics[height=4.52cm]{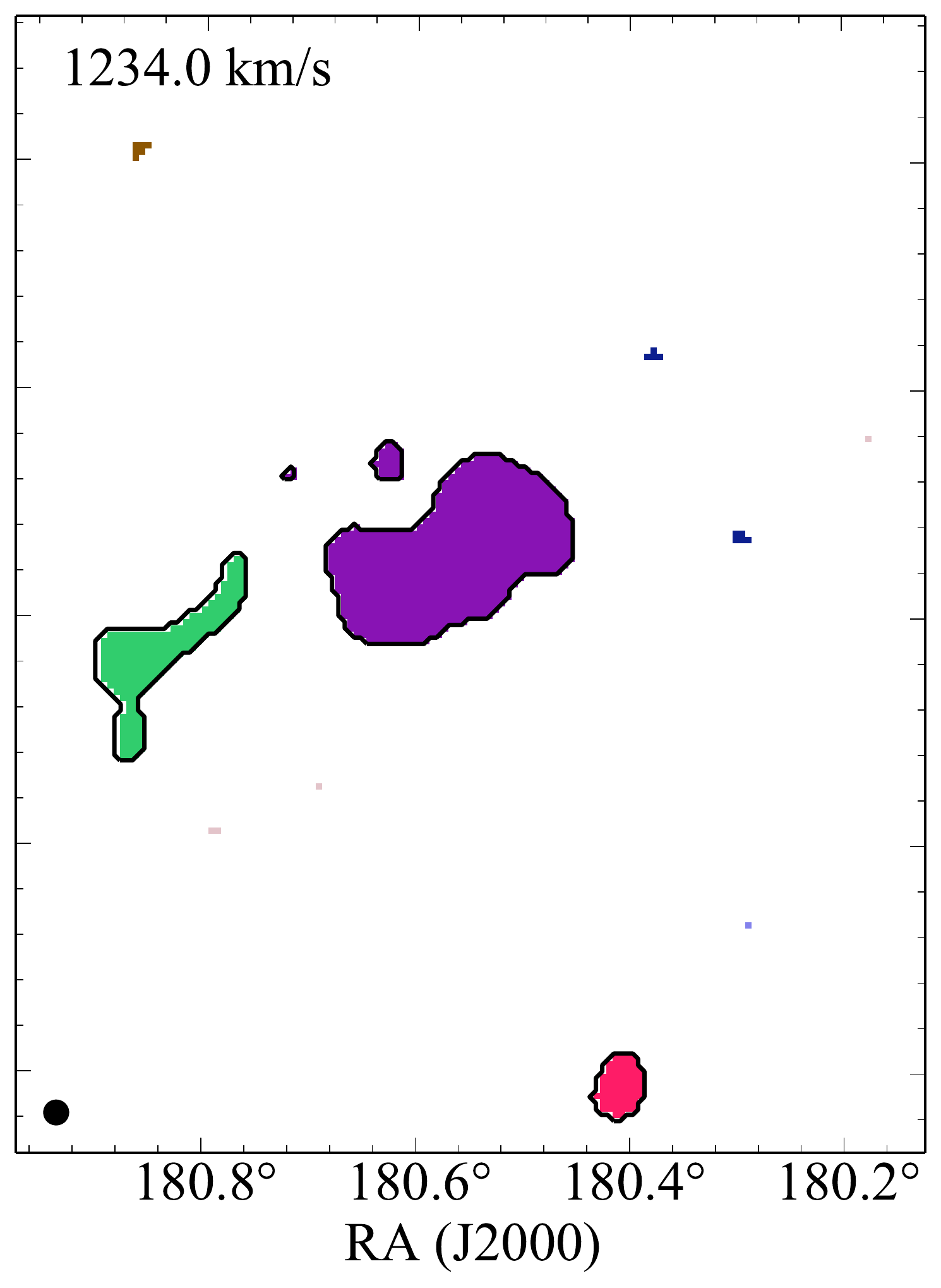}
\includegraphics[height=4.52cm]{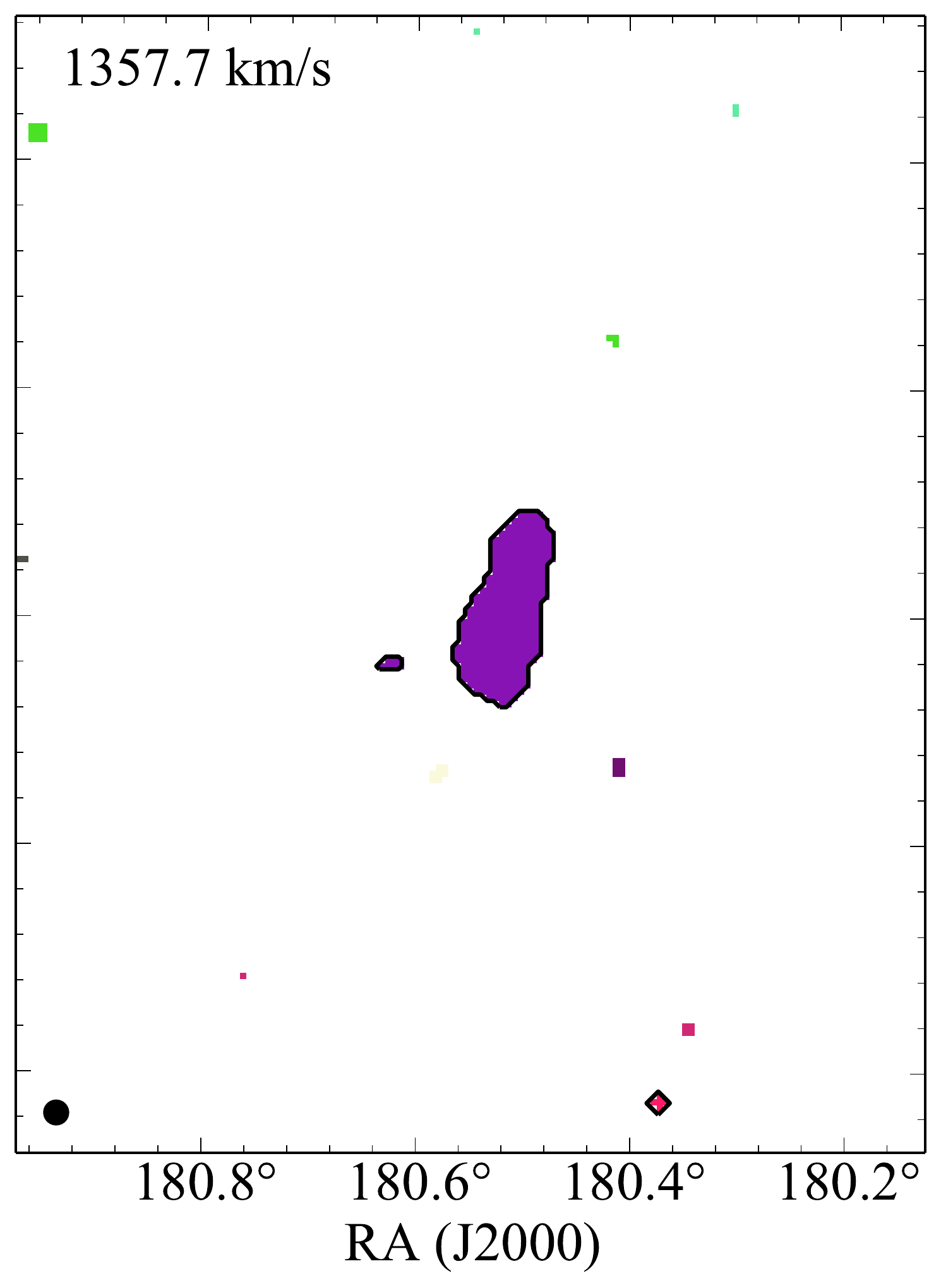}
\includegraphics[height=4.52cm]{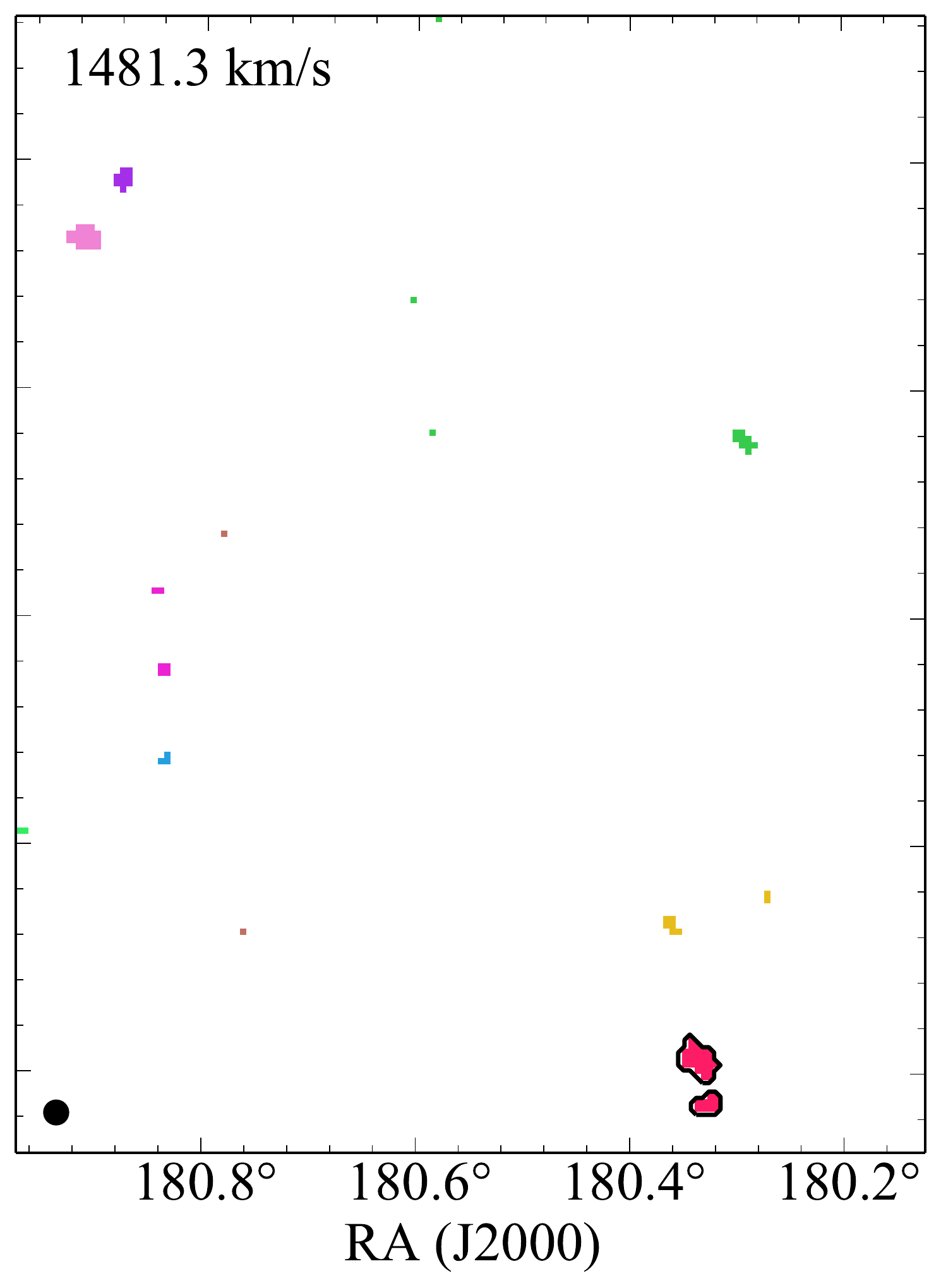}
\includegraphics[height=4.52cm]{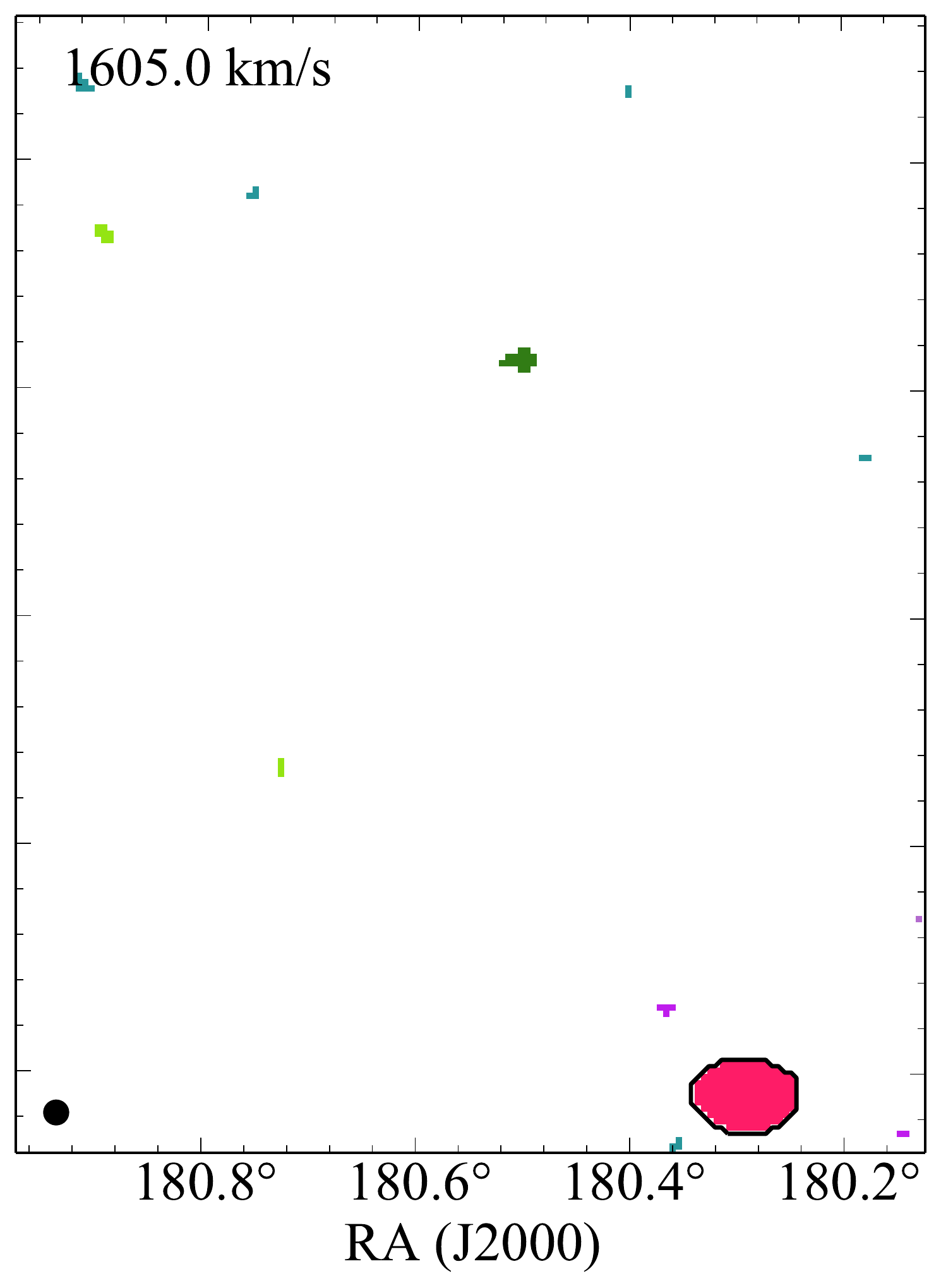}
\caption{Illustration of the detection of signal and identification of individual sources in \sofia. \it Top panels. \rm Channel maps extracted from the data cube shown in Fig. \ref{fig:paperfiguredavide}. The line-of-sight velocity of each channel is indicated in the top-left corner (note that these are not adjacent channels in the original cube). The beam is shown in the bottom-left corner. Black contours show regions included in the binary mask (Sec. \ref{sec:det}). \it Bottom panels. \rm Same channel maps as in the top panels but now showing the individual objects formed on the basis of the binary mask (Sec. \ref{sec:obj}). We show only objects with positive total flux. Each object is indicated with a different random colour. Black contours indicate the four objects whose reliability is higher than 99 per cent (Sec. \ref{sec:rel}).}
\label{fig:paperfigure1}
\end{figure*}

All the above algorithms return a binary mask of detected voxels (and any additional source-finding algorithm could be added to \sofia\ as long as they satisfy this condition). As an example, the top panels of Fig. \ref{fig:paperfigure1} show five channels extracted from the data cube in Fig. \ref{fig:paperfiguredavide} and, with black contours, the regions included in the binary mask. In this case the S+C finder was employed using 12 different smoothing kernels. The relatively low adopted threshold ($3.5 \sigma$) results in a number of noise peaks being included in the mask. We come back to this point in Sec. \ref{sec:rel}.

\subsection{Merging detected voxels into sources}
\label{sec:obj}

The aforementioned binary mask is the basis for identifying individual sources or objects. In \sofia, this computationally expensive operation is performed using the C++ implementation of the \cite{lutz1980} one-pass algorithm by \cite{jurek2012}, combined with a sparse representation of 3D objects. We refer to \cite{jurek2012} for details on this implementation. Here it is sufficient to say that this algorithm produces the same result as a friends-of-friends method with linking element equal to an elliptic cylinder. Users can specify the cylinder size. This step of \sofia\ also returns basic source parameters such as total flux, peak flux (both normalised by the noise level) and size.

The bottom panels of Fig. \ref{fig:paperfigure1} show the objects created from the binary mask using a merging cylinder with a radius of 3 pixels and a height of 7 channels (we show only objects with positive total flux). These panels show four real detections as well as a number of positive noise-peak objects. It is worth highlighting the successful detection of a faint, extended \hi\ tail east of the brightest galaxy (second panel from the left). This detection is made possible by the fact that \sofia\ looks for emission on multiple scales. Furthermore, \sofia\ correctly identifies as a single source the resolved, edge-on galaxy located in the southern part of the cube (visible in all panels but the first) despite the low level emission at channels close to the systemic velocity.

\subsection{Reliability and rejection of false detections}
\label{sec:rel}

All detection algorithms listed above require users to specify a detection threshold. The closer this threshold is to the noise, the more noise peaks will be included in the resulting binary mask. Some of these noise peaks may be identified as separate objects if they are sufficiently far from a real object (see bottom panels of Fig. \ref{fig:paperfigure1}). \sofia\ offers two ways of removing these false detections from the final output.

The first method is a simple size filter and is based on the fact that all real detections are at least as large as the data cube's resolution. In practice, users can specify the minimum acceptable source size along each axis of the cube independently. The downside of this method is that it may potentially remove relatively bright but unresolved sources from the final object mask.

The second method is illustrated by \cite{serra2012b} and estimates the reliability of individual objects by comparing the distribution of positive and negative sources (i.e., sources with positive and negative total flux, respectively) in parameter space. The simple idea is that the distribution of positive and negative noise peaks should be identical while positive, real detections should not have a negative counterpart in parameter space. It is based on the assumptions that the noise is symmetric and that real sources have positive total flux (i.e., absorption line sources have been masked).

Within \sofia, the reliability can be calculated following the run of any source-finding algorithm chosen by the user as long as both positive and negative noise peaks are included in the binary mask, and after the detected voxels are merged into sources (Sec. \ref{sec:obj}). The reliability calculation also requires that a sufficient number of negative noise peaks is included in the mask such that their distribution in parameter space can be studied meaningfully. Users can select to produce diagnostic plots on the reliability calculation similar to those shown in \cite{serra2012b}. The black contours in the bottom panels of Fig. \ref{fig:paperfigure1} highlight objects whose reliability is higher than 99 per cent. 

In summary, users can decide to run \sofia\ with a high detection threshold, resulting in a reliable but possibly incomplete catalogue of detections; but they can also decide to dig deeper into the noise using a lower threshold, and successively remove false detections. In the latter case, a reliability value can be returned for all positive detections.

\subsection{Mask optimization}

\sofia\ measures the parameters of all sources (e.g., total flux, size, line width) considering only voxels included in the mask cube. However, experience shows that masks can miss the faint, outer edge of objects, in particular if obtained with a high detection threshold. This would introduce systematic effects in the measured parameters (e.g., the total flux would be underestimated; see \citealt{westmeier2012}). To prevent this, \sofia\ offers two mask optimization methods which modify the object mask cube by growing the masks which define individual objects. In both methods, the mask is grown independently for each object.

The first method is mostly appropriate for sources that are unresolved on the sky or, if resolved, face-on and symmetric. It starts by fitting an ellipse to the moment-0 image of the object. The ellipse is then used as a mask for all velocity channels occupied by the object -- i.e., the initial mask, which generally has an arbitrary 3D shape, is converted into an elliptic cylinder. Finally, the size of the ellipse is increased until a maximum in total flux is reached (a similar method is described by \citealt{barden2012} in the context of 2D imaging).

\begin{figure*}
\includegraphics[height=3.78cm]{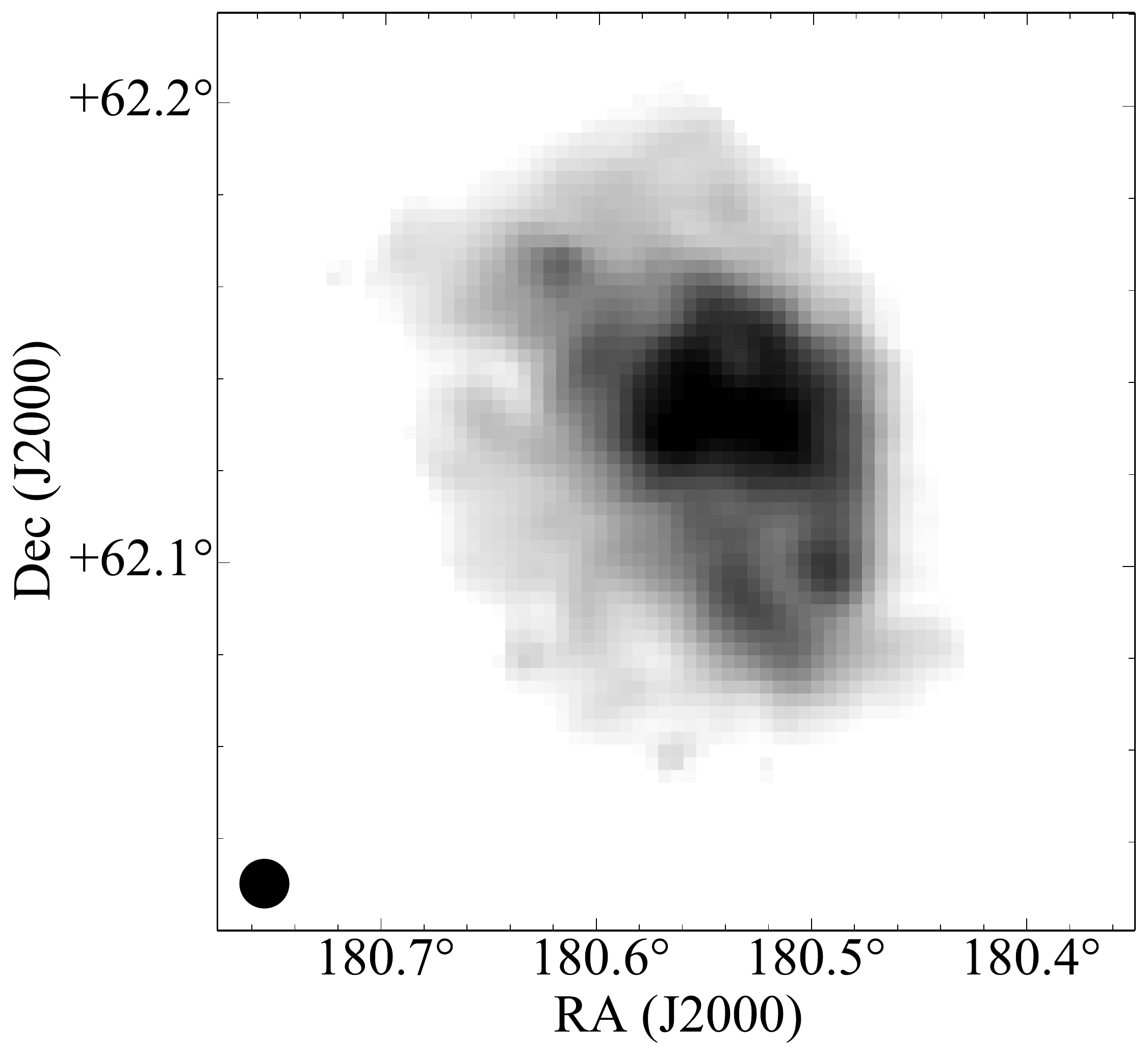}
\includegraphics[height=3.78cm]{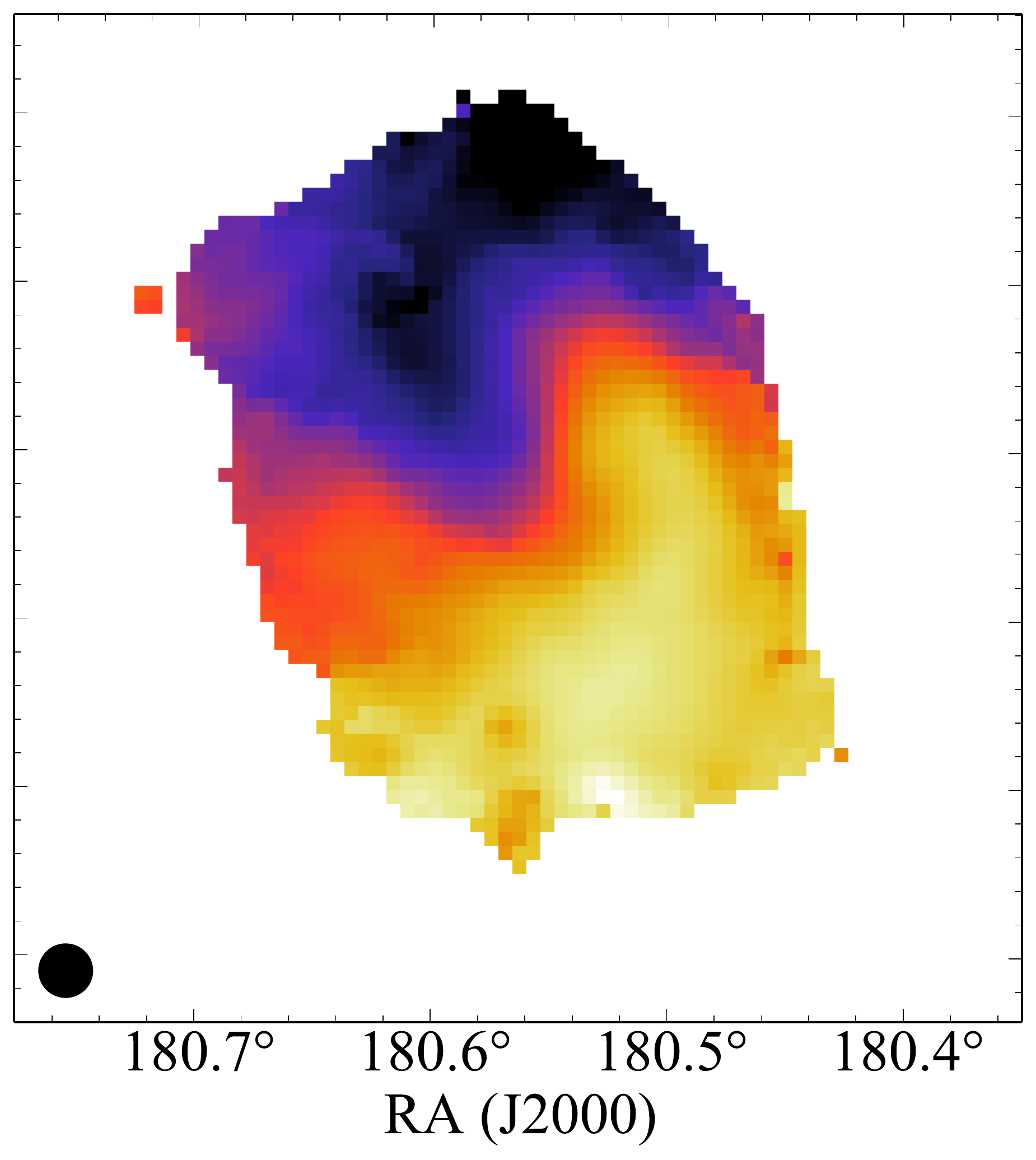}
\includegraphics[height=3.942cm]{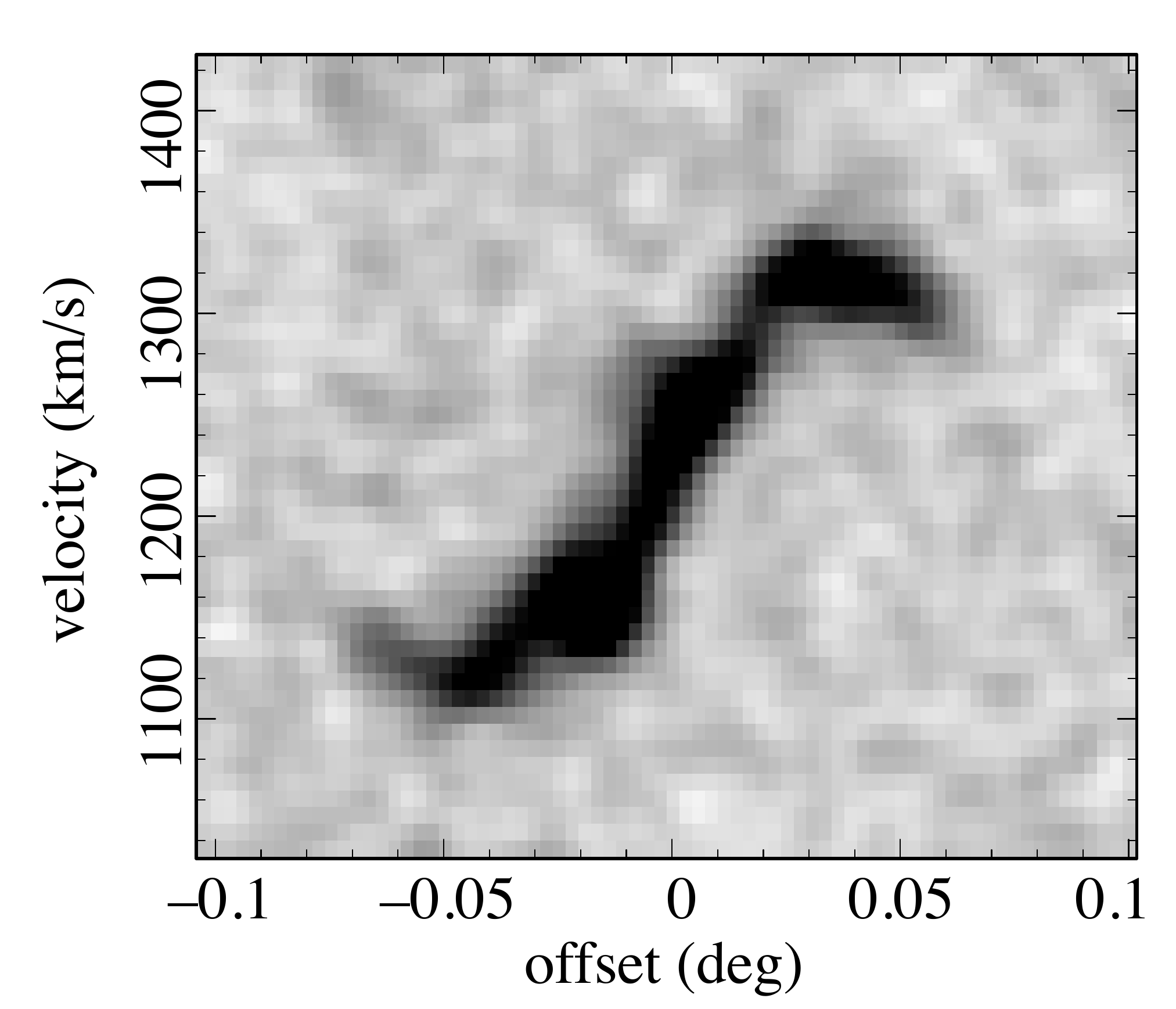}
\includegraphics[height=3.85cm]{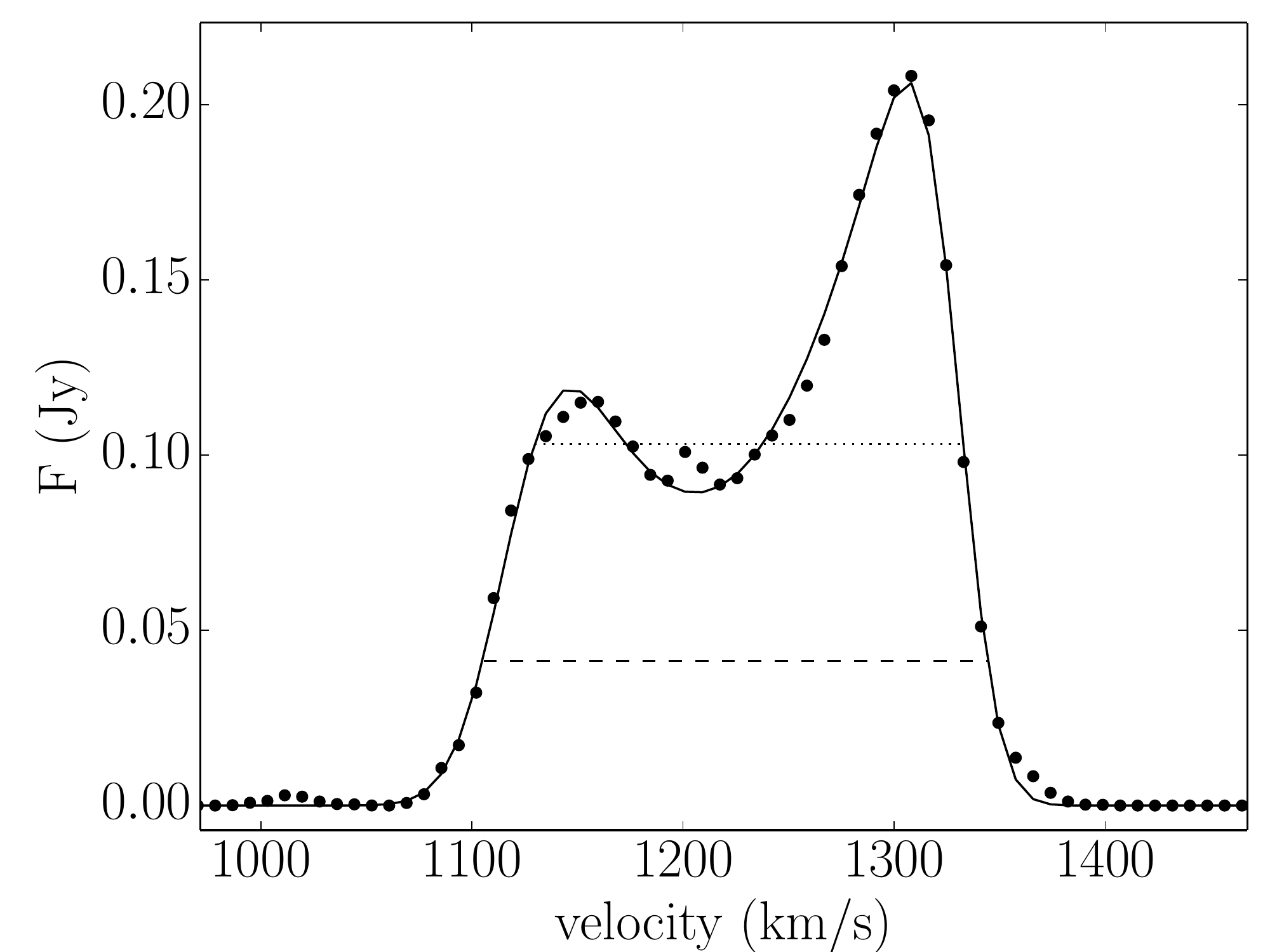}
\caption{Data products for the brightest galaxy in the cube shown in Fig. \ref{fig:paperfiguredavide}. From left to right: moment-0 image; moment-1 velocity field; position-velocity diagram along the morphological major axis ($\mathrm{PA}=32^\circ$); and integrated spectrum (filled circles) with the best-fitting busy function overlaid (solid line). In the latter panel the dotted and dashed lines indicate the line widths $W_{50}$ and $W_{20}$ estimated at 50 and 20 per cent of the peak flux, respectively, on the basis of the busy function fit.}
\label{fig:paperfigure2}
\end{figure*}

The above method should in principle be applied only to sources which fill most of the cylindrical mask in all channels (see above), while for objects with a more complex 3D structure it can result in a decrease of the integrated signal-to-noise ratio. For this reason we provide a second mask growth method. This consists in performing a binary dilation of the initial mask along the two spatial axes of the data cube using a 2D dilation structuring element whose shape approximates a circle. The size of the structuring element is increased iteratively until the total flux converges (i.e., until the relative flux growth between successive iterations is lower than a threshold specified by the user). This method preserves the 3D shape of the initial mask. In addition to growth along the two spatial axes, this algorithm can also grow all masks by a fixed number of channels (selected by the user) along the frequency axis.

\subsection{Source parameterization}

As mentioned in Sec. \ref{sec:obj}, basic source parameters are measured when detected voxels are merged into objects. These can be used to estimate the reliability of each source and reject false detections in parameter space. After mask optimization \sofia\ re-computes those parameters and measures additional ones. These include: position (both geometric and centre of mass); total flux; minimum and maximum voxel value; size and bounding region along each axis; line width measured using different methods (including the one proposed by \citealt{courtois2009}); results of an ellipse fit to the moment-0 image; results of a busy function fit to the integrated spectrum (for a description of the busy function see \citealt{westmeier2014}). These parameters are provided both in a ``raw'' format (i.e., coordinates in pixel units, fluxes in data units, line-width in channels) as well as converted into more useful units (e.g., WCS coordinates and standard flux and velocity units). Some of these parameterization steps are optional and implementation of additional parameters is straightforward within the code.

\subsection{Output products}

Users can decide what output \sofia\ should produce. Available output products include:

\begin{itemize}
\item Catalog of objects and their parameters, both in ASCII and VO-compliant XML format.
\item Final object mask.
\item Moment 0 and 1 images of the sky area covered by the full data cube determined from the data within the mask.
\item Cut-out data cubes containing individual objects as well as their corresponding mask, moment 0, 1 and 2 images, integrated spectrum and position-velocity diagram along the morphological major axis. An example of these products is shown in Fig. \ref{fig:paperfigure2}.
\end{itemize}
In future releases it will be possible to produce these products for just a subset of the detections by selecting a region of interest in source parameter space.

This output is designed to not only give useful information about the detected sources but also to enable further, higher-level analysis by the user. For example, the cut-out cubes of individual objects and the corresponding masks could be used to measure additional source parameters not included in \sofia\ or to produce Gauss-Hermite velocity fields to enable kinematical studies.

\subsection{Performance of \sofia}
\label{sec:memory}

In the current implementation of \sofia\ the entire input data cube (or the selected sub-cube; see Fig. \ref{fig:flowchart}) is loaded into memory and processed on a single core. Additional cubes will also need to be stored in memory at various stages of processing, such as the weights cube, the binary or object mask cube, and a smoothed version of the data cube if required by the source-finding algorithm being used (e.g., S+C), plus a potentially large array of source parameters. It is therefore interesting to discuss how the memory requirement and execution time of \sofia\ vary with cube size. For this purpose we make use of two cubes. The smaller cube is the one used  for illustration purpose in this paper (Figs. \ref{fig:paperfiguredavide} and \ref{fig:paperfigure1}). It has 360 pixels along both spatial axes and 150 channels along the frequency axis, resulting in a file size of 78 MB. The second cube is the one used for the source-finding tests of \cite{serra2012b} and \cite{westmeier2012}. It too has 360 pixels along both spatial axes but consists of 1464 channels along the frequency axis. Therefore, its size is $\sim10$ times that of the first cube.

We process the two cubes with identical settings employing a representative combination of the algorithms described in this paper: noise normalisation along the frequency axis; S+C source finding with 12 smoothing kernels; merging of detected voxels into sources; calculation of reliability and removal of unreliable sources; optimization of the mask of individual objects using the dilation method; source parameterization including busy function fit; creation of output products for the cubes as a whole and for the individual detections; creation of ASCII and XML catalogues. The two runs are carried out on a machine running Linux Mint 17 with a memory of 16 GB and a 2.9 GHz Intel Core processor.

\begin{figure*}
\includegraphics[width=18cm]{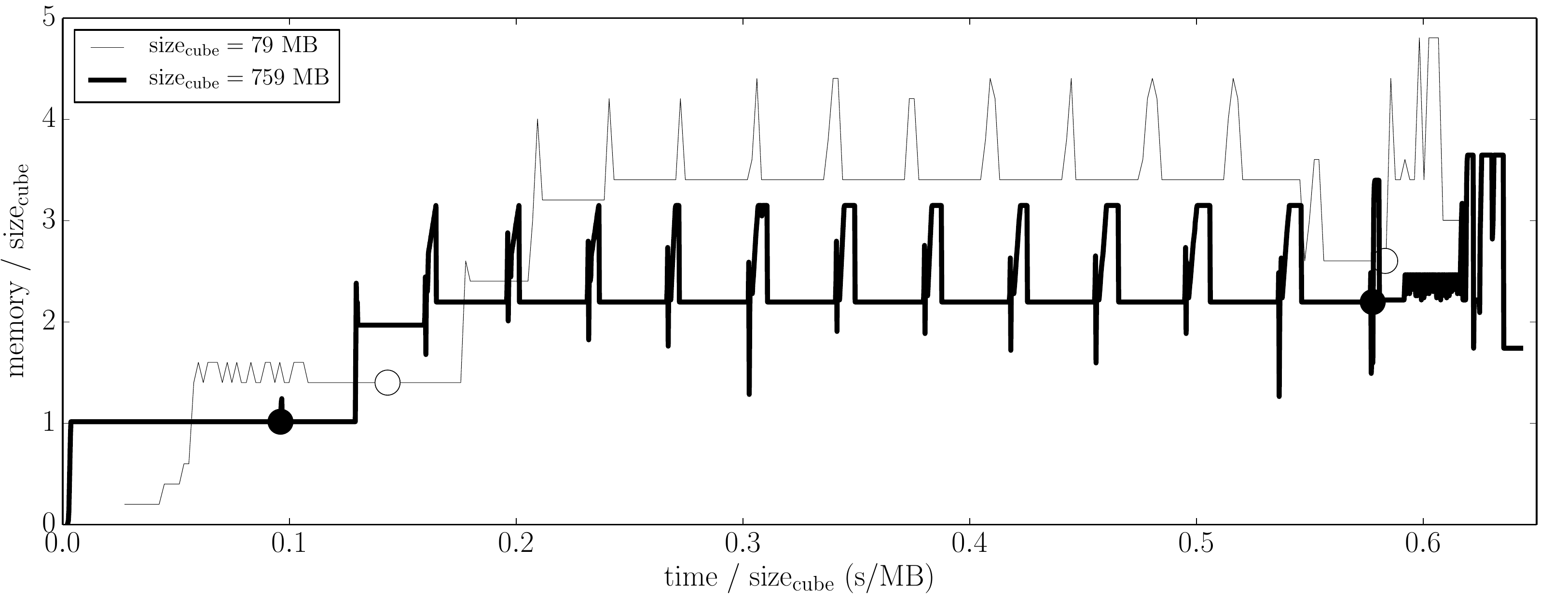}
\caption{Memory usage as a function of time for two runs of \sofia\ on two cubes whose sizes differ by a factor $\sim10$ (see legend). As explained in the text, both runs include the use of the S+C source finder, and open and filled black circles indicate the beginning and the end of the S+C execution for the two cubes, respectively. See Sec. \ref{sec:memory} for a more detailed discussion of the time taken by other algorithms during the two \sofia\ runs.}
\label{fig:memory}
\end{figure*}

Figure \ref{fig:memory} shows the memory usage of \sofia\ as a function of time for the two cubes. Both axes of the plot are normalised by the cube size. The time behaviour of the two curves appears very similar, indicating that the execution time scales approximately linearly with cube size within the range explored here. The memory offset between the two curves is due to the loading into memory of a number of libraries used by \sofia. These come with a memory overhead of the order of a few tens of megabytes, which is more noticeable in the case of a smaller data cube. For data cubes much larger than this overhead the memory usage is between 2 and 3 times the size of the cube, with occasional peaks between 3 and 4 times the cube size.

Figure \ref{fig:memory} allows us to investigate the memory and processing time taken by the various algorithms. In this case, most of the time is taken by the S+C finder. The beginning and end of its execution are marked by open and filled black circles for the small and large data cube, respectively.  The peaks in memory usage of S+C correspond to the smoothing operations, while the plateaus in between peaks correspond to the noise level calculation. Each additional filter would contribute another $\sim0.05$ s/MB to the execution time in this case. The noise calculation appears to be particularly time consuming. In this case, it is carried out using the aforementioned Gaussian fit to the negative side of flux histogram (Sec. \ref{sec:cubes}). This calculation uses the full cube, and an obvious way to increase its speed would be to calculate the noise level on a sub-cube. This option may become available in the future.

The time before the beginning of the S+C finder in Fig. \ref{fig:memory} is taken by the noise normalisation along the frequency axis and by an initial measurement of the noise level in the normalised cube. The time after S+C is taken by all other algorithms listed above, and these are typically much faster. The memory peak right at the end of S+C corresponds to the merging of voxels into sources. The height of this peak depends on the number of sources detected. This is followed by the calculation of the reliability and the rejection of unreliable sources, which are both relatively inexpensive in terms of memory but can be time consuming. The final memory peaks correspond to the creation of moment images.

\subsection{Source-finding based on a catalogue of 3D coordinates}

The above discussion makes it clear that \sofia\ is currently not able to process arbitrarily large data cubes but is limited by the memory of the system on which it is run. This problem is partially alleviated by the fact that \sofia\ is able to limit the processing to a sub-cube whose boundaries are specified by the user. Therefore, users could choose to run \sofia\ multiple times on sufficiently small portions of a large input cube, obtaining individual output products for each of them. They could then combine these products, creating for example a single mask or catalogue. In the future we may be able to offer such breaking up of a large input cube into sub-cubes -- and the creation of final data products for the full cube -- as a processing mode fully  integrated with the other modules of \sofia.

In this context, a useful feature already available in \sofia\ is that it allows users to search for emission in any number of small sub-cubes centred at a set of 3D coordinates within an arbitrarily large data cube. For example, in an era of large \hi\ and optical redshift surveys, this mode could be used to look for emission in a large \hi\ cube at the location of galaxies included in an optical spectroscopic catalogue.

This mode is fully integrated in \sofia\ and interested users need to simply provide the input data cube and a catalogue of 3D coordinates. \sofia\ will process the various positions sequentially, each time loading into memory only the sub-cube of interest. The 3D size of the sub-cubes can be set by the user and is the same for all positions. Users can also request the creation of a single output catalogue of sources, which is generated by merging the catalogues obtained at each position.

\subsection{Comparison to other source finders}

A number of established software packages for the reduction and analysis of interferometric data allow some source finding to be carried out on data cubes (e.g., \texttt{GIPSY}, \texttt{Miriad}). However, this approach requires users to develop custom codes which make use of (and are limited by) the tasks available within those general-purpose packages. In contrast, the more specialised \sofia\ offers a wide range of  ready-to-use source-finding algorithms, which are already integrated with one another and can be combined in a flexible way to produce a variety of output products.

The other 3D application for spectral line data which shares some of these characteristics is \duchamp\ \citep{whiting2012a}. This application detects sources using a simple threshold method (similar to the one described in Sec. \ref{sec:det}) and then grows them using a secondary threshold. This algorithm differs from those available in \sofia\ and, in this respect, the two packages could be seen as complementary (\citealt{popping2012} shows that \duchamp\ has the best performance for unresolved sources but does not reach the completeness of S+C for resolved sources). With respect to memory requirements, \duchamp\ is similar to \sofia\ in that it loads and processes in one core the full input data cube. Therefore, it too is limited by the memory of the system on which it is run.

A significant advantage of \sofia\ compared to \duchamp\ is that it offers a larger number of algorithms for both source finding and parameterization. This includes the S+C and CNHI finders, the 2D-1D wavelet de-noising (denoising is available in \duchamp\ but it uses isotropic wavelets, which are not ideal for spectral line sources whose size along the spectral axis is decoupled from the size along the two spatial axes), the calculation of the reliability of individual detections, the mask optimization by binary dilation, the possibility of searching for signal on the basis of an input catalogue of 3D coordinates and the busy function fit. The creation of cubelets and PV diagrams for individual detections is also not included in \duchamp\ but is available in \texttt{Selavy}, a source finder built upon \duchamp\ for distributed processing of large cubes \citep{whiting2012b}. While future development may reduce the difference between \duchamp/\texttt{Selavy} and our package, all above methods are at the moment unique to \sofia.

Finally, it is worth mentioning that \sofia\ does not offer at the moment a full analysis of the sources' morphology. For example, a group of nearby detected voxels is merged into a single source regardless of the size of the source and only based on the merging element chosen by the user (Sec. \ref{sec:obj}). This means that no information is given about whether the source, which could be very large, is composed of distinct and easily recognisable components. Different and more specialised source finders are able to provide such characterisation (e.g., \texttt{Clumpfind} by \citealt{williams1994}; \texttt{BLOBCAT} by \citealt{hales2012}). We note however that the object mask cube returned by \sofia\ could be used as a starting point for further morphological analysis of the detections.

\section{Summary}

We provide a high-level description of \sofia, a flexible source finder for 3D spectral line data. \sofia\ puts together for the first time in a single package a number of new source-finding and parameterization algorithms developed in preparation of upcoming \hi\ surveys with ASKAP (WALLABY, DINGO) and APERTIF. It is, however, designed to enable the use of these new algorithms on any data cube independent of emission line or telescope used.

We describe the various methods and algorithms available in \sofia\ as well as planned developments. One key advantage of \sofia\ is that it allows users to search for spectral line signal on multiple scales on the sky and in frequency (using. e.g., the S+C finder or the 2D-1D wavelet filter), which is crucial to detect and parameterize 3D sources in a complete and reliable way. Furthermore, within \sofia\ it is possible to take into account noise level variations across the cube and the presence of errors and artefacts. Moreover, \sofia\ is able to estimate the reliability of individual detections, which should be particularly useful for surveys expected to detect a large number of sources. It can also produce a variety of output products, including moment images, cut-out cubes and images, integrated spectra and catalogues of source parameters. Finally, \sofia\ is able to search for line emission in arbitrarily large data cubes on the basis of a catalogue of 3D coordinates. Most of these methods are not available in other source finders and are currently unique to \sofia.

We provide a few visual examples of how \sofia\ works including a view of the dedicated  graphical user interface. We describe the available parameterization and the wide range of output products, which include mask cubes, moment images, position-velocity diagrams and busy function spectral fits of individual sources. This output is designed to both provide a useful description of the sources as well as facilitate subsequent analysis.

We highlight the modularity of \sofia, which allows users to optimize the source-finding and parameterization strategy for the data and sources of interest. This modularity also enables future expansions of \sofia\ to include new source-finding and parameterisation algorithms.

\sofia\ is publicly available at the website indicated in Sec. \ref{sec:intro}  together with technical information on how to use the software. Software updates, improvements and bug fixes are posted regularly at this webpage. \sofia\ is registered at the Astrophysics Source Code Library with ID ascl:1412.001.

\section*{Acknowledgments}

The authors acknowledge financial support from a Research Collaboration Award of the University of Western Australia. TvdH and NG were supported by the European Research Council under under the European Union's Seventh Framework Programme (FP/2007-2013) / ERC Grant Agreement nr. 291531. LF acknowledges support by the Deutsche Forschungsgemeinschaft (DFG) under grant numbers KE757/7-1, KE757/7-2, KE757/7-3 and KE757/9-1. LF is a member of the International Max Planck Research School (IMPRS) for Astronomy and Astrophysics at the Universities of Bonn and Cologne.

\bibliographystyle{mn2e}
\bibliography{sofiapaper}

\label{lastpage}

\end{document}